\documentclass[cameraready]{Interspeech}
\title{EmoSURA: Towards Accurate Evaluation of Detailed and Long-Context Emotional Speech Captions}

\author[affiliation={1,2}, orcid=0000-0002-8803-9414]{Xin}{Jing}
\author[affiliation={1,2}, orcid=0000-0001-8338-617X]{Andreas}{Triantafyllopoulos}
\author[affiliation={1,2}, orcid=0000-0001-9372-3133]{Jiadong}{Wang}
\author[affiliation={1,2,3}]{Shahin}{Amiriparian}
\author[affiliation={3}]{Jun}{Luo}
\author[affiliation={1,2,4},orcid=0000-0002-6478-8699]{Björn}{Schuller}
\address{
    $^1$ CHI -- Chair of Health Informatics, TUM University Hospital, Munich, Germany, \\
    $^2$ MCML - Munich Center for Machine Learning, Munich, Germany\\
    $^3$ Huawei, Netherlands, \\
    $^4$ GLAM, Imperial College London, UK
}

\email{xin.jing@tum.de}

 %BS: if space allows, add a 5th one for SEO

\usepackage{comment}
\usepackage[capitalize]{cleveref}
\usepackage{graphicx}
\usepackage{subcaption}
\usepackage{booktabs} 
\usepackage{multirow}
\usepackage{amsmath} 
\usepackage{xcolor} 
\usepackage{tabularx}

\begin{document}

\maketitle

\begin{abstract}
Recent advancements in speech captioning models have enabled the generation of rich, fine-grained captions for emotional speech. However, the evaluation of such captions remains a critical bottleneck: traditional N-gram metrics \textbf{fail} to capture semantic nuances, while LLM judges often suffer from reasoning inconsistency and context-collapse when processing long-form descriptions. In this work, we propose 
%BS: deleted "the" 
EmoSURA, a novel evaluation framework that shifts the paradigm from holistic scoring to atomic verification. EmoSURA decomposes complex captions into \textit{Atomic Perceptual Units}, which are self-contained statements regarding vocal or emotional attributes, and employs an audio-grounded verification mechanism to validate each unit against the raw speech signal. Furthermore, we address the scarcity of standardized evaluation resources by introducing \textbf{SURABench}, a carefully balanced and stratified benchmark. 
Our experiments show that EmoSURA achieves a positive correlation with human judgments, offering a more reliable assessment for long-form captions compared to traditional metrics, which demonstrated negative correlations due to their sensitivity to caption length. 
% Code and resources will be available.
\end{abstract}

\section{Introduction}

Driven by recent advances in large-scale audio–language models (ALMs), emotional speech captioning can now generate highly fluent, long, and detailed natural language descriptions of a speaker's vocal characteristics, emotional state, and prosodic style \cite{Jing25-Qwen3, Huang25-SAI, Xu24-SSE}. 
Nevertheless, we are still lacking a thorough understanding of what makes a generated caption \emph{appropriate} according for human listeners -- an understanding which is critical for guiding future model development \cite{Triantafyllopoulos25-CAF, Schuller26-ACH}.

Evaluating emotional speech captions is non-trivial due to the mismatch between the nature of the task and existing evaluation metrics \cite{Sarto25-ICE}. Traditional n-gram–based metrics focus on surface-level lexical overlap and are ill-suited for evaluating free-form, perceptually grounded descriptions \cite{Ye25-PWW}. Semantic similarity metrics \cite{Dixit25-MLAF, Zhang19-BET, Zhao19-MTG}  
partially address this limitation by operating in learned 
%BS: everywhere was American English, so I changed the few British ones to it...
embedding spaces, but they remain sensitive to text length and often fail to adequately assess long, information-dense captions. 
More recently, large language models (LLMs) have been adopted as judges due to their ability to process complex natural language \cite{Guo25-BAB, Wang25-STG}. However, when presented directly with long captions containing rich and fine-grained emotional details, LLM-as-a-judge approaches are prone to information loss and inconsistent reasoning, reducing their overall reliability \cite{Mohsin25-OTF, Barkett25-RIE}. To mitigate the complexity of evaluating such long texts, some works attempt to decompose captions into discrete labels or keywords prior to LLM scoring \cite{Ando24-FSC, Spataru24-KWT, Lightman23-LVS}. However, decoupling the text from the source audio fundamentally restricts the model's flexibility, rendering it incapable of directly grounding emotional descriptions in the actual acoustic signals.

\begin{figure*}[t]
  \centering
  \includegraphics[width=0.8\linewidth]{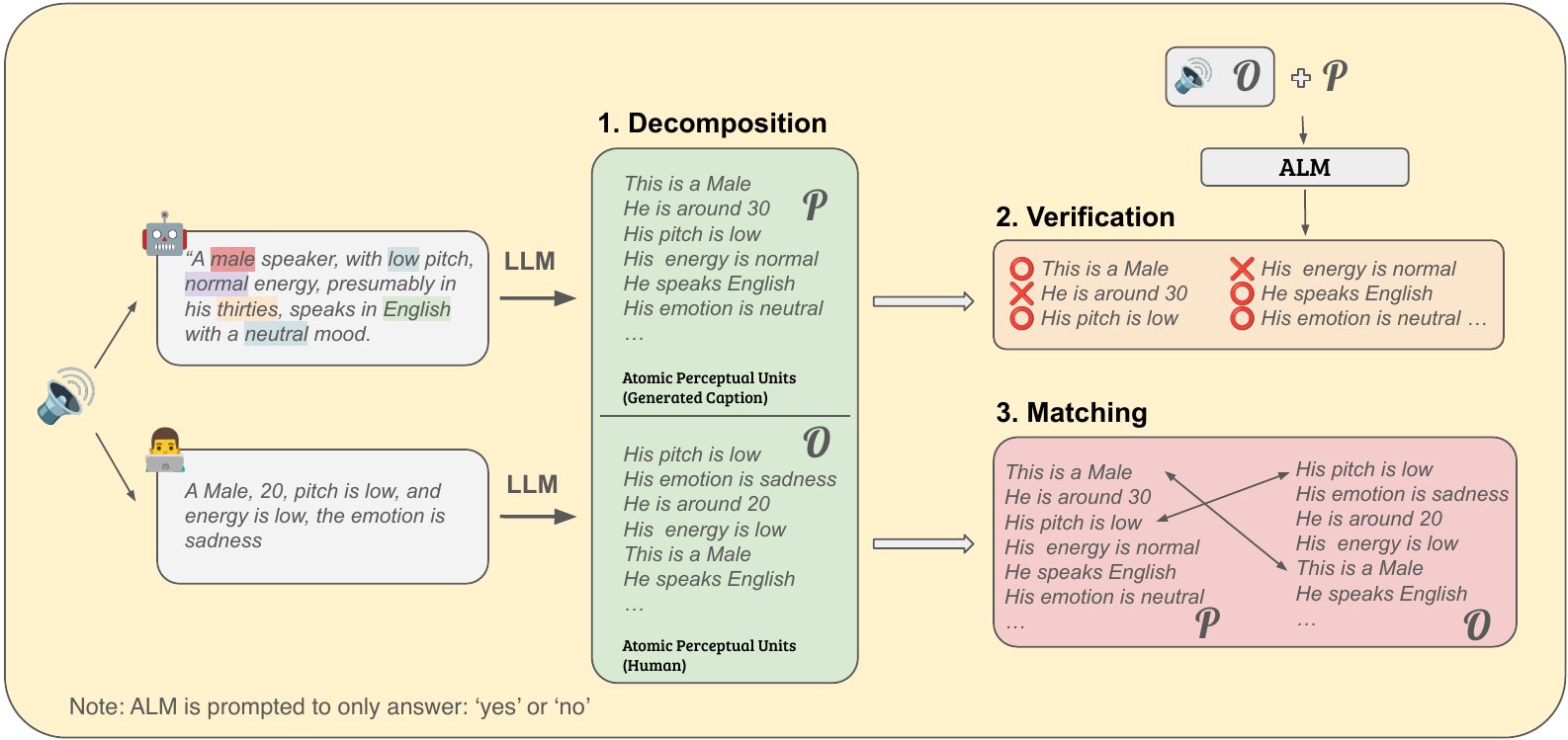}
  \caption{The framework of EmoSURA. It consists of three steps: (1) Decomposition of captions into Atomic Perceptual Units (APUs) using LLMs; (2) Verification of generated APUs against the raw audio using an ALM; and (3) Matching generated APUs with benchmark to assess comprehensiveness.}
  \label{fig:sys}
\end{figure*}

%BS: Figure 1: Add "." after "the emotion is sadness" (lower left corner)
%BS: Figure 1: "Note: ALM only is asked..." sounds grammatically wrong - please check.
%BS: Why in Figure 1 "This is a Male" and not "This is a male"? (2x)

To address these limitations, we propose  \textbf{Emo}tional \textbf{S}peech \textbf{U}nderstanding \textbf{R}ating Score (\textbf{EmoSURA}), a structured evaluation framework for emotional speech captions. The core idea of EmoSURA is to decompose a caption into a set of minimal atomic perceptual units (APUs), where each unit is expressed as a complete, self-contained \textit{sentence} describing a single vocal or emotional attribute. Rather than evaluating captions holistically, EmoSURA assesses each information unit independently by querying an audio–language model to determine whether the corresponding perceptual evidence is present in the original speech data. 
To mitigate hallucination and cascading errors, unit-level evaluation is formulated as a binary decision task, where the model is limited to \textit{yes/no} judgments regarding the presence of each perceptual unit in the original audio. Such constrained decision schemes have been shown to yield more stable and consistent evaluations in prior LLM-based assessment studies  \cite{Pham24-HHP}. The final score reflects both the perceptual validity of individual units and the alignment between candidate captions and reference captions.
This ``decompose-and-evaluate'' strategy offers several advantages. By operating on APUs, EmoSURA improves interpretability and enables fine-grained error analysis, as each evaluation step produces explicit intermediate decisions, which facilitate error tracing and model refinement. The framework is flexible, enabling new evaluation dimensions to be introduced without redesigning the metric.

The contributions of this work are threefold:

\begin{itemize} 
\item We propose \textbf{EmoSURA}, a novel fine-grained evaluation framework that decomposes captions into APUs and leverages audio-grounded verification. 

\item We construct \textbf{SURABench}, a balanced and stratified benchmark dataset, providing a standardized resource for reproducible reference-based evaluation. 

\item We demonstrate through comprehensive experiments that EmoSURA achieves \textbf{state-of-the-art correlation performance with human judgments} compared to traditional N-gram and embedding-based metrics, proving its robustness in detecting hallucinations and evaluating fine-grained paralinguistic details. 
\end{itemize}

The remainder of this paper is organized as follows.
We describe the proposed EmoSURA framework and SURABench in \cref{sec:met} and \cref{sec:data}.
In \cref{sec:exp}, we present the experimental setup, ablation studies, and results, followed by a detailed analysis.
Finally, in \cref{sec:con}, we summarize our findings and discuss future directions.

\section{EmoSURA}
\label{sec:met}

As illustrated in ~\cref{fig:sys}, EmoSURA is a modular evaluation framework designed to assess the alignment between emotional speech and captions. Rather than producing a single monolithic score, EmoSURA deconstructs the evaluation process into three interpretable stages: \textit{Decomposition}, \textit{Verification}, and \textit{Matching}. Each step targets a common failure in emotional speech captioning: sentence-level ambiguity, acoustic hallucinations, and content under-specification.
EmoSURA operates at the granularity of APUs: minimal, self-contained semantic assertions extracted from captions. This atomic approach enables fine-grained analysis and explicit reasoning regarding hallucinated or missing content.

\subsection{Step 1: Atomic Decomposition}
The first step of EmoSURA decomposes complex captions into Atomic Primitive Units (APUs). Crucially, each APU is formulated as a \textbf{standalone declarative statement}. This format is essential because only complete propositions possess a well-defined truth value, enabling the subsequent model to perform robust binary (Yes/No) verification. This design effectively mitigates the ambiguity and semantic entanglement inherent in coarse-grained sentence-level evaluation \cite{Jing24-FFE}.
Given a caption set $\mathcal{C}$, we employ \texttt{Qwen2.5-7B-Instruct} as a decomposition engine. Let $\mathcal{P} = \{p_1, p_2, \ldots, p_N\}$ denote the set of APUs extracted from a generated caption $\mathcal{C}_{gen}$, and $\mathcal{O} = \{o_1, o_2, \ldots, o_M\}$ denote the corresponding set extracted from the human-annotated reference caption $\mathcal{C}_{ref}$. The model is prompted to parse each caption into atomic statements, where each unit encodes a single subject-predicate-object relation or attribute-level fact. 

\subsection{Step 2: Audio-Grounded Verification}
A key limitation of traditional text-based evaluation metrics lies in their inability to penalize hallucinations, i.e., affective or acoustic descriptions not grounded in the underlying audio signal. To address this limitation, EmoSURA introduces an audio-grounded verification stage that explicitly evaluates the factual plausibility of generated APUs with respect to the acoustic evidence.
For each generated unit $p_i \in \mathcal{P}$, we employ \texttt{Qwen2-Audio-7B-Instruct} as an audio-grounded judge. The ALM jointly processes the raw speech signal $\mathcal{A}$ and the textual unit $p_i$, and is explicitly prompted to perform a \textit{binary} entailment judgment (\textit{Yes/No}) indicating whether the audio supports the statement. This formulation follows established LLM-as-a-judge paradigms widely adopted in recent evaluation literature, accounting for the tendency of longer descriptions to exhibit higher hallucination rates due to error accumulation \cite{Ye25-PWW, Pham24-HHP, Zhang23-HLM}.
To ensure robustness, the verification process is designed to be conservative, prioritizing the rejection of falsifiable affective or acoustic hallucinations over the acceptance of ambiguous or partially supported descriptions. Formally, the verification function is defined as:
$V(p_i \mid \mathcal{A}) = \text{ALM}(\mathcal{A}, p_i) \in \{\text{Yes}, \text{No}\}.$
Accordingly, the set of hallucination-free units supported by the audio is given by:
$\mathcal{P}_{true} = \{ p_i \in \mathcal{P} \mid V(p_i \mid \mathcal{A}) = \text{Yes} \}.$
This verification stage yields a precision-oriented score that reflects factual correctness with respect to the audio signal: $s_p = |\mathcal{P}_{true}|/|\mathcal{P}|$.

\subsection{Step 3: Semantic Matching}
While audio-grounded verification assesses factual correctness, it does not capture whether the generated caption sufficiently covers the salient content present in the reference description. To assess completeness and mitigate omission errors, EmoSURA evaluates semantic recall at the APU level.
We employ \texttt{Qwen2.5-7B-Instruct} to perform semantic alignment between generated units and reference units. For each reference unit $o_j \in \mathcal{O}$, the model determines whether it is semantically entailed or matched by at least one generated unit $p_i \in \mathcal{P}$. The subset of matched reference units is denoted as $\mathcal{Q} \subseteq \mathcal{O}$.
Importantly, in addition to reference-aligned units, we explicitly account for verified generated units that convey correct but non-reference information. This prevents valid, audio-supported descriptions from being penalized as false negatives. The recall-oriented score is therefore defined as:
\begin{equation}
s_r = \frac{|\mathcal{Q}| + |\mathcal{P}_{true} \setminus \mathcal{Q}|}{|\mathcal{O}| + |\mathcal{P}_{true} \setminus \mathcal{Q}|}.
\end{equation}

This formulation rewards both faithful coverage of reference content and the inclusion of additional, verifiable descriptive details.

\subsection{Final Scoring}
Based on the precision-oriented score $s_p$ and recall-oriented score $s_r$, we compute an overall F1 score to balance factual correctness and content coverage:
\begin{equation}
s_f = 2 \cdot \frac{s_p \cdot s_r}{s_p + s_r}.
\end{equation}

In addition, to explicitly assess descriptive richness, we compute a descriptive F1 score $s_f'$, following the same formulation but restricted to descriptive APUs only. The final EmoSURA score is defined as:
\begin{equation}
\mathcal{F} = \frac{1}{2} \left( s_f + s_f' \right).
\end{equation}

\section{SURABench}
\label{sec:data}
\begin{figure}
    \centering
    \includegraphics[width=0.7\linewidth, trim={0cm 0cm 0cm 0.7cm}, clip]{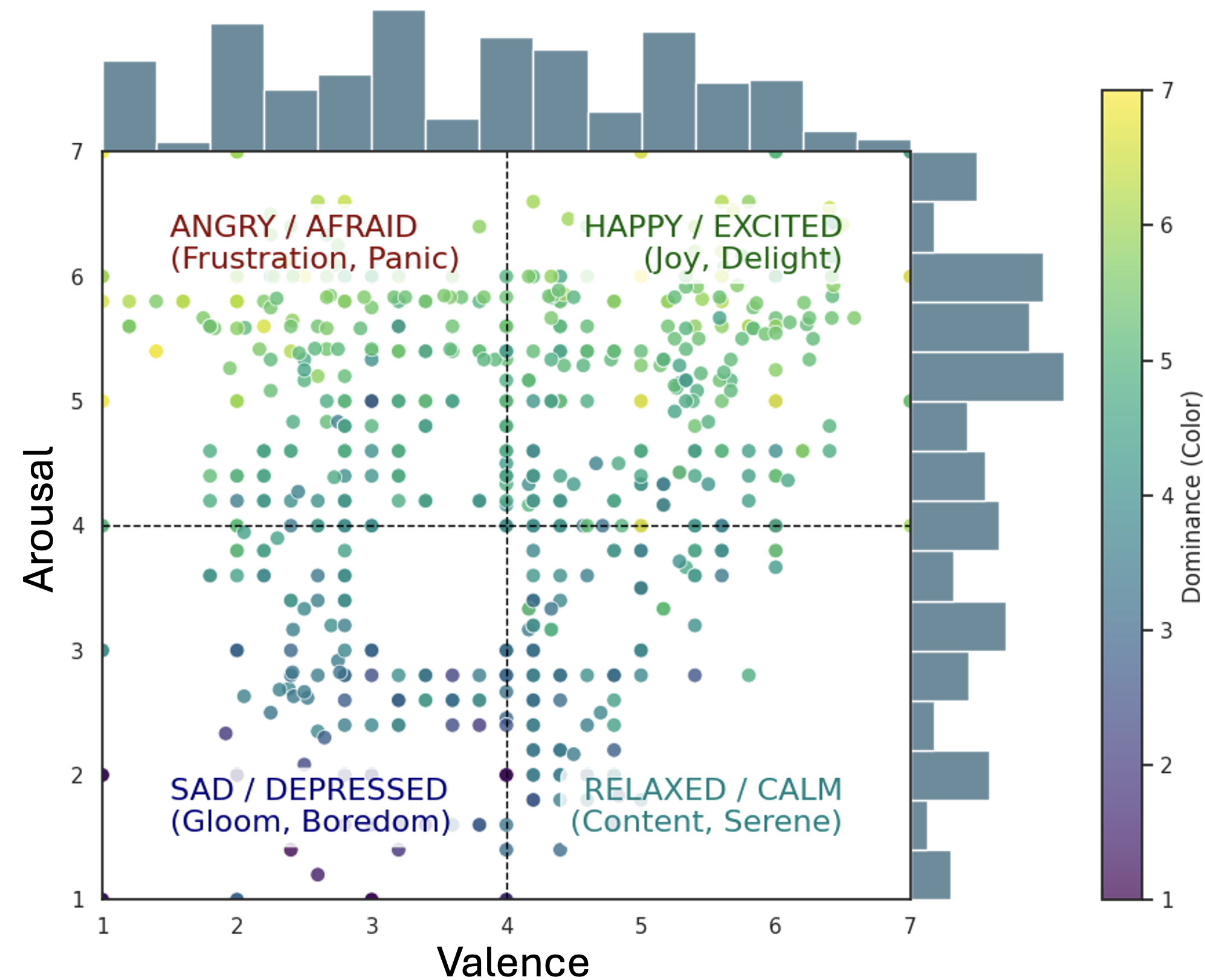}
    \caption{The emotional distribution of SURABench in the Valence-Arousal space. Point colors represent Dominance. The marginal histograms (top and right) demonstrate the uniformity of the dataset.} %BS: where is a HEAT MAP to explain the colors?! You MUST add it!
    \label{fig:bench}
\end{figure}

For a high-quality evaluation foundation, we construct \textbf{SURABench}, a balanced evaluation benchmark derived from the MSP-Podcast v1.11 \textit{Test1} split \cite{Busso25-TMC}. 
A three-stage curation process was applied to guarantee acoustic suitability, label reliability, and distributional balance.
We firstly excluded segments that are too short ($< 3s$) to convey complete emotional semantics or too long ($> 8s$) to allow for stable caption generation.
Then, we enforced a consensus constraint, keeping only utterances where the standard deviations of both valence and arousal ratings were $\le 1.5$. This step removes ambiguous samples with low inter-annotator agreement.
Finally, we implemented a stratified grid sampling strategy to mitigate the severe class imbalance. We discretized the Valence-Arousal space (scaled $1\text{-}7$) into a $10 \times 10$ grid. From each bin, we selected up to 15 samples, prioritizing those with the highest annotation consensus (lowest variance). As showed in ~\cref{fig:bench}, this sampling mechanism ensures uniform coverage across all four semantic quadrants while reducing the over-representation of neutral speech. The resulting dataset comprises\textbf{ 1,018 }utterances with broad emotional coverage.

To generate high-fidelity captions, we implemented a widely-adopted hybrid annotation pipeline that synergizes acoustic feature extraction with human-guided LLM generation \cite{Bai2025-AAE, Mei24-WAC, Sun24-AAL}.
First, we followed ParaCLAP \cite{Jing24-PTA, Jing25-EET} to extract a set of paralinguistic features to serve as objective evidence, including pitch, pitch variation, loudness, jitter, shimmer, and speech tempo.
Then, expert annotators manually drafted ``gold-standard'' descriptions for a representative subset of the data based on the ground-truth emotion labels. These human-authored captions establish the desired standard for descriptive granularity and syntactic style. Finally, we prompted GPT-4.1 with ``gold-standard'' descriptions as few-shot samples to generate captions for the entire benchmark. 

\begin{figure*}[]
  \centering
  \includegraphics[width=0.9\linewidth]{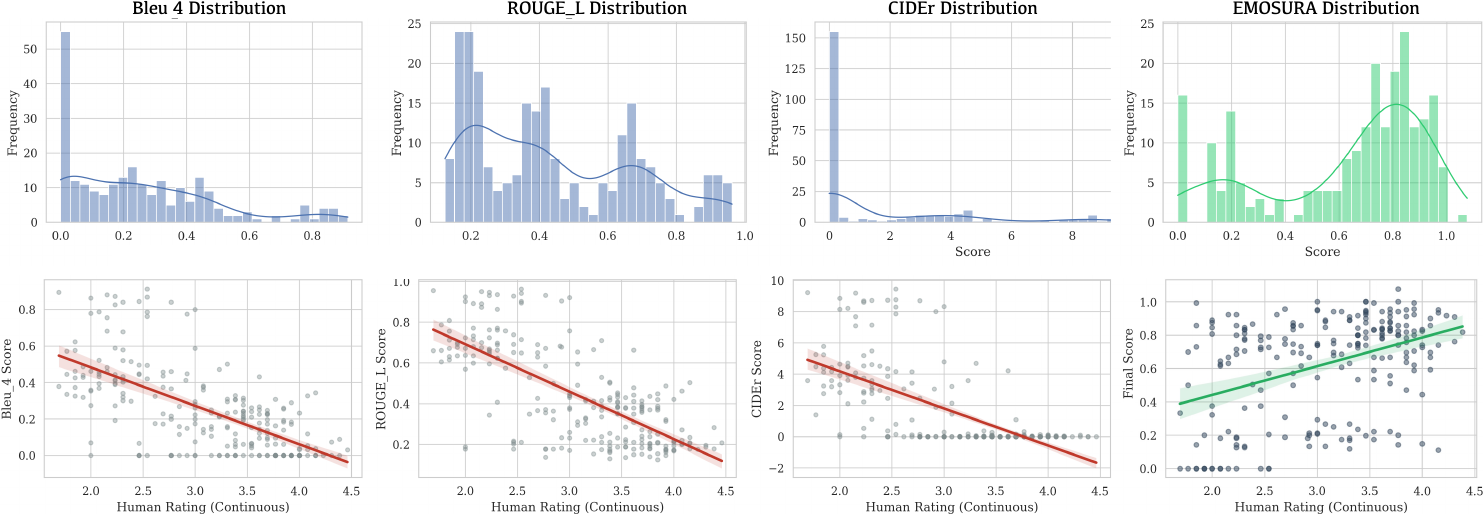}
  \caption{Distribution plots (top) and scatter plots against human ratings (bottom) for baseline metrics versus EMOSURA. Unlike Bleu 4, ROUGE\_L, and CIDEr, which show negative correlations (red lines), EMOSURA demonstrates a positive linear relationship (green line) with human ground truth, indicating higher reliability.}
  \vspace{-4mm}
  \label{fig:metric_nlp}
\end{figure*}
\section{Experiments \& Analysis}
\label{sec:exp}
\subsection{Subjective test}
We performed a subjective Mean Opinion Score (MOS) evaluation involving 14 participants (6 males, 8 females, including 6 audio experts) 
%BS: you MUST add more on dempgraphics! AGE! And possibly native language, background knowledge, etc. !
on 320 audio-caption pairs. To ensure representativeness, these pairs were stratified across the Valence--Arousal circumplex and speaker gender. Evaluators were instructed to listen to the audio and then rated samples on a 5-point Likert scale.
%BS: Something could be wrong here - was the audio already rated (this is what is written!) or did they have to rate (which I assume) - then, you MUST reword! 
The samples were integrated in four specific caption categories: (i) Ground Truth (GT); 
%BS: Ground Truth and gold standard are DIFFERENT THINGS! I think you write something wrong here - please CHECK! 
(ii) Sabotaged Captions, where specific factual details were corrupted to test error detection; (iii) Unconstrained Qwen-Omni, representing long-form, free-style generation; and (iv) Refined Qwen-Omni, representing concise, schema-adherent outputs.

%BS: Figure 2 is WRONG!!! Arousal is NOT (!) Intensity - please DELETE "(Intensity)"!!!

As shown in \cref{tab:main_leaderboard}, all rule-based metrics exhibit negative correlations with human judgments across all three correlation coefficients. As seen at the top row of \cref{fig:metric_nlp}, BLEU-4 and CIDEr distributions are heavily zero-inflated and skewed towards the lower bound, indicating that for rich, descriptive captions, the strict n-gram matching requirement of traditional metrics fails to assign meaningful scores to semantically correct but lexically diverse descriptions. Meanwhile, model-based metrics show mixed behavior. SPICE remains negatively correlated, while MACE achieves moderate positive correlations. In contrast, EmoSURA demonstrates consistent positive alignment with human ratings. Compared to MACE, EmoSURA achieves higher rank correlations, showing improved consistency in ranking samples according to human preferences.

\begin{table}[t]
\centering
\caption{Correlation between automated evaluation metrics and human ratings. Performance is measured using  Pearson correlation coefficient (PCC), Kendall's rank correlation coefficient (Kd $\tau$) and sample-wise $\tau$ (Sp $\tau$). All p-values $<$ 0.001.} 
%BS: WHICH significance test? Make sure it is named somewhere (perhaps already is...)! If space allows, explain "Kd" and "Sp" in the caption.
\label{tab:main_leaderboard}
\renewcommand{\arraystretch}{1.2}
\begin{tabular}{l|ccc}
\toprule
\textbf{Metric} & \textbf{PCC ($\rho$) $\uparrow$} & \textbf{Kd $\tau$ $\uparrow$} & \textbf{Sp $\tau$ $\uparrow$} \\
\midrule
\multicolumn{4}{l}{\textit{\textbf{Rule-Based Metrics}}} \\
\midrule
BLEU-4 \cite{Papineni02-BAMF}   & -0.6419 & -0.4494 & -0.6916 \\
ROUGE-L \cite{Lin24-VOPF}  & -0.7017 & -0.4606 & -0.6916 \\
METEOR \cite{Banerjee05-MAAM}   & -0.5813 & -0.4541 & -0.6559 \\
CIDEr \cite{Vedantam15-CCID}    & -0.6640 & -0.3732 & -0.6175 \\
SPIDER \cite{Liu17-IIC}   & -0.6679 & -0.4481 & -0.7019 \\
\midrule
\multicolumn{4}{l}{\textit{\textbf{Model-Based Metrics}}} \\
\midrule
SPICE \cite{Anderson16-SSPI}    & -0.5728 & -0.3874 & -0.6240 \\
MACE \cite{Dixit25-MLAF}     & 0.4283  & 0.2619  & 0.3709 \\
\midrule
\multicolumn{4}{l}{\textit{\textbf{Proposed Method}}} \\
\midrule
\textbf{EmoSURA (Ours)} & \textbf{0.4391} & \textbf{0.3277} & \textbf{0.4480} \\
\bottomrule
\end{tabular}
\end{table}

Our statistical analysis also reveals a fundamental distributional shift between ground truth references and model-generated captions. While the target references maintain a controlled length ($\mu = 459$ chars, $\sigma \approx 60$), SOTA models like Qwen-omni exhibit a
%BS: p-value? test method? ADD!!!
tendency towards verbosity, producing captions with an average length of $684$ characters, which is approximately 1.5 times longer than the references. This length discrepancy is fatal for precision-based metrics. Since n-gram metrics treat non-overlapping tokens as ``insertion errors'', the additional 200+ characters generated by the model -- even if they contain valid, hallucination-free visual details -- are heavily penalized. The high variance in generation length ($\sigma \approx 280$, with extremes reaching 1,318 characters) further exacerbates this issue, causing the erratic scoring behavior observed in \cref{fig:metric_nlp}. Meanwhile, we analyzed the instruction adherence of the Audio-LLM judge. Across the entire test set, we observed a low format failure rate of 5.61\%, where the model failed to produce a valid binary \textit{(`Yes'/`No')} token. This demonstrates that Qwen-Omni effectively adheres to the constrained verification schema without significant generation collapse. However, despite outperforming baselines, the PCC ($\rho \approx 0.44$) between EmoSURA and human performance indicates that a substantial portion of the variance in human judgment remains unexplained, suggesting that EmoSURA has not yet fully captured the nuances of human cognitive appraisal. 

\subsection{Perturbation Test}
We construct a controlled perturbation evaluation set based on SURABench by systematically modifying GT captions to introduce attribute-level inconsistencies with the corresponding audio. Specifically, we define an acoustic hallucination as any caption attribute that contradicts the annotated acoustic evidence (e.g., speaker gender, emotional state, vocal events, or low-level acoustic features).
Perturbations are applied in a controlled manner to alter a single semantic attribute per sample while maintaining acoustic-semantic consistency; for instance, a demographic swap from male to female was deliberately paired with corresponding modifications in pitch and vocal texture descriptions \cite{Alahmari25-LLM}. The modified samples are distributed across three predefined hallucination categories: Emotion Flip, Vocal Event Fabrication, and Acoustic Feature Swap. Please refer to Supplementary Sec. 2 for a comprehensive definition.

\begin{table}[htbp]
\centering
\caption{Detection performance of \textbf{EmoSURA} across different hallucination categories. The system demonstrates exceptional sensitivity to demographic and acoustic factual errors.}
\label{tab:emosura_detection}
\begin{tabularx}{\columnwidth}{X|ccc}
\toprule
\textbf{Perturbation} & \textbf{Injected} & \textbf{Detected} & \textbf{Rate (\%)} \\
\midrule
\textbf{Acoustics Feat.} & \textbf{120} & \textbf{112} & \textbf{93.33} \\
\quad \textit{-- Gender} & 40 & 39 & 97.50 \\
\quad \textit{-- Pitch, Tempo, Vol., etc.} & 80 & 73 & 91.25 \\
\midrule
\textbf{Emotion} & \textbf{40} & \textbf{33} & \textbf{82.50} \\
\midrule
\textbf{Vocal Event} & \textbf{40} & \textbf{24} & \textbf{60.00} \\
\bottomrule
\end{tabularx}
\end{table}

As detailed in \cref{tab:emosura_detection}, the detection performance of EmoSURA exhibits a clear degradation from low-level acoustic features to high-level semantic behaviors. The system demonstrates exceptional robustness against physical acoustic and emotion perturbations. This indicates a highly reliable cross-modal alignment between the audio and text modalities for fundamental, frame-level attributes like gender, pitch, and tempo. Therefore, EmoSURA could leverage its strong acoustic sensitivity to detect emotional polarity flips.

Conversely, the detection of fabricated vocal events (e.g., hallucinating singing or sobbing over normal speech) forms a distinct performance bottleneck, dropping to 60.00\%. 
% Unlike static demographic traits or emotional states, 
We think that vocal events demand complex, long-term temporal modeling and higher-level semantic abstraction. This performance gap may suggest that while EmoSURA acts as a highly precise `acoustic fact-checker,' it struggles to capture the temporal dynamics of complex vocalizations.

\section{Conclusion}
\label{sec:con}

%BS: conclusion in past tense - I changed...
This work addressed the evaluation bottleneck in emotional speech captioning by introducing a granular verification metric and a standardized benchmark. The proposed framework, EmoSURA, shifts from holistic text scoring to an atomic verification approach. By decomposing captions and verifying individual claims against the raw audio, the method explicitly targets the issues of hallucination and sentence-level ambiguity common in long-form generation. We also developed SURABench, a spontaneous speech corpora with long and detailed captions. Our experimental analysis indicates that EmoSURA aligns better with human perception than existing rule-based and embedding-based metrics, which tend to penalize the verbosity of modern generative models. Future work will focus on utilizing EmoSURA's feedback in reinforcement learning to directly optimize the factual consistency of captioning models.

% \section{Acknowledgments}

% % {\color{blue}Acknowledgments should be included only in the camera-ready version, not in the version submitted for review. For regular papers, pages 5 and 6, and for long papers, pages 9 and 10, are reserved exclusively for acknowledgments, disclosures of the use of generative AI tools, and references. No other content may appear on these pages. Any appendices must be contained within the first four pages for regular papers and within the first eight pages for long papers.

% % Acknowledgments and references may begin on an earlier page if space permits.}

% \ifcameraready
%      The Interspeech 2026 organizers
% \else
%      The authors
% \fi
% would like to thank ISCA and the organizing committees of past Interspeech conferences for their help and for kindly providing the previous version of this template.

% {\color{blue}

% \newpage

\bibliographystyle{IEEEtran}
\bibliography{mybib}
\newpage
\textbf{Generative AI Use Disclosure}
The authors used a generative AI tool to assist with language editing and manuscript polishing. The tool was used solely to improve clarity, grammar, and readability. All scientific content, analyses, interpretations, and conclusions were developed and verified by the authors.

\end{document}

\section{Prompts for EmoSURA}

% --- Prompt 1 ---
\begin{promptbox}{Prompt 1: Decomposition}
\#\#\# Persona\\
You are an expert linguistic analyst. Your specialty is deconstructing descriptive text into verifiable, primitive information units that characterize vocal, speech, and emotional patterns.

\vspace{0.5em}
\#\#\# Task\\
Analyze the provided input text. For each vocal, speech, or emotional characteristic you identify, extract it as a primitive information unit. Your output must be a JSON list of objects, where each object represents a single, verifiable attribute found in the text.

\vspace{0.5em}
\#\#\# Rules
\begin{enumerate}[leftmargin=*, noitemsep, topsep=0pt]
    \item Read the input text carefully to identify keywords, phrases, and pronouns that provide direct evidence about vocal and emotional characteristics.
    \item For each piece of evidence, create a corresponding JSON object in the output list.
    \item The \code{"fact"} key must contain a complete, standalone sentence describing the characteristic.
    \item The \code{"evidence"} key must contain the exact quote from the input text that supports the fact.
    \item If no evidence is present for a particular attribute, do not create an object for it.
    \item Your final output must be only the JSON list and nothing else.
\end{enumerate}

\vspace{0.5em}
\#\#\# Output Schema and Allowed Terms\\
Each object in the JSON list must conform to the following structure:
\begin{itemize}[leftmargin=*, noitemsep, topsep=0pt]
    \item \code{fact}: (string) A complete sentence describing the attribute.
    \item \code{attribute}: (string) Must be one of: \code{"pitch"}, \code{"rate"}, \code{"volume"}, \code{"gender"}, \code{"emotion"}.
    \item \code{value}: (string) The specific term for the attribute. Allowed terms are:
    \begin{itemize}[leftmargin=*, noitemsep]
        \item For \code{"pitch"}: \code{"low"}, \code{"normal"}, \code{"high"}.
        \item For \code{"rate"}: \code{"slow"}, \code{"normal"}, \code{"fast"}.
    \end{itemize}
    \item \code{evidence}: (string) The exact substring from the input text.
\end{itemize}

\vspace{0.5em}
\#\#\# Examples

\textit{Example 1}\\
Input: "His voice is deep..."\\
Output:
\begin{verbatim}
[
    {
        "fact": "The speaker's gender is male.",
        "attribute": "gender",
        "value": "male",
        "evidence": "His"
    }
]
\end{verbatim}

\vspace{0.5em}
\#\#\# Input Text
\end{promptbox}

\newpage

% --- Prompt 2 ---
\begin{promptbox}{Prompt 2: Audio Verification}
You are an expert audio verifier.

\vspace{0.5em}
\#\#\# Task\\
You will be given a "Fact", an audio file, and a set of "Ground Truth" text. Your task is to determine if the "Fact" is correct.

\vspace{0.5em}
\#\#\# Rules
\begin{enumerate}[leftmargin=*, noitemsep, topsep=0pt]
    \item \textbf{Audio is Primary:} Your primary judgment must be based \textit{only} on what is heard in the audio.
    \item \textbf{GT is Secondary:} The GT text is a reference.
    \item \textbf{Conflict Resolution Rule:} The audio is the ultimate source of truth.
\end{enumerate}

\vspace{0.5em}
\textbf{Output}
\begin{itemize}[noitemsep, topsep=0pt]
    \item 1 $\rightarrow$ the fact is correct
    \item 0 $\rightarrow$ the fact is incorrect
\end{itemize}

> Ground Truth: \{GT\}

Input:\\
> Fact:
\end{promptbox}

% --- Prompt 3 ---
\begin{promptbox}{Prompt 3: Semantic Matching}
You are now a audio-linguistic expert in matching two set of primitive information units generated from two captions.

The set of primitive information units is represented as a list of dict \code{[\{"fact": [UNIT], "identifier": [ID]\}...]} within JSON format. In addition, each primitive information unit in the oracle set would be accompanied with a unique "id".

To match primitive information units... (Instructions omitted for brevity)...

\vspace{0.5em}
\textbf{IMPORTANT}: Please consider each primitive information unit in the set individually. You should only output the matching results formatted as:
\code{[\{"fact": [UNIT], "identifier": [ID], "matched\_oracle\_id": [ORACLE ID]\}...]}

The "identifier" is optional. For key named \code{"matched\_oracle\_id"}, the value should be the corresponding "id". If not matched, set \code{"matched\_oracle\_id"} to "None".

\vspace{0.5em}
> > > Oracle Set: \{GT\}\\
> > > Set of Primitive information units:
\end{promptbox}

\section{Case Study: Qualitative Analysis of Sabotaged Captions}
\label{sec:case_study}

To rigorously evaluate the system's robustness, we manually crafted adversarial negative samples (sabotaged captions) by systematically perturbing the ground-truth annotations. As outlined in our taxonomy, these perturbations span four distinct categories of increasing complexity. Table \ref{tab:case_study} presents qualitative examples for each category, highlighting the specific lexical substitutions used to inject semantic and acoustic hallucinations.

\begin{itemize}
    \item \textbf{Type A: Emotion Polarity Flip.} Targets the affective dimension by reversing the emotional valence and arousal while maintaining the underlying factual events.
    \item \textbf{Type B: Attribute Swap.} Injects demographic factual errors by altering the speaker's biological identity (e.g., gender) and naturally adjusting the associated acoustic properties (e.g., pitch).
    \item \textbf{Type C: Event Fabrication.} Fundamentally alters the temporal and behavioral nature of the audio event, replacing normal speech with stylized vocalizations such as singing or sobbing.
    \item \textbf{Type D: Mixed Hallucination.} The most severe perturbation, representing a complete semantic collapse where gender, emotion, vocal behavior, and physical acoustics are inverted simultaneously.
\end{itemize}

\begin{table*}[htbp]
\centering
\caption{Case examples of the four hallucination categories. The injected hallucinations and altered acoustic descriptions are highlighted in \textcolor{red}{\textbf{bold red}} to illustrate the contrast against the ground-truth (GT) captions.}
\label{tab:case_study}
\renewcommand{\arraystretch}{1.4}
\begin{tabularx}{\textwidth}{@{} c >{\hsize=0.5\hsize}X >{\hsize=1.5\hsize}X @{}}
\toprule
\textbf{Type} & \textbf{Version} & \textbf{Caption Text} \\ 
\midrule

\multirow{2}{*}{\textbf{A}} & \textit{GT Caption} & A male speaker delivers the line... in a \textbf{bold, vibrant} voice brimming with \textbf{confident enthusiasm}. His tone is \textbf{commanding and energetic}, projecting \textbf{authority} while radiating \textbf{positivity}... \\ \cline{2-3}
& \textit{Sabotaged} & A male speaker delivers the line... in a \textcolor{red}{\textbf{weak, shaky}} voice brimming with \textcolor{red}{\textbf{anxious dread}}. His tone is \textcolor{red}{\textbf{fearful and hesitant}}, projecting \textcolor{red}{\textbf{vulnerability}} while radiating \textcolor{red}{\textbf{negativity}}... \\ 
\midrule

\multirow{2}{*}{\textbf{B}} & \textit{GT Caption} & A \textbf{male} speaker introduces \textbf{himself}... in a \textbf{gruff, commanding} manner. Despite the \textbf{low, rumbling pitch}... his voice carries an \textbf{intense, forceful} edge... The texture is \textbf{rough and raspy}... \\ \cline{2-3}
& \textit{Sabotaged} & A \textcolor{red}{\textbf{female}} speaker introduces \textcolor{red}{\textbf{herself}}... in a \textcolor{red}{\textbf{soft, gentle}} manner. Despite the \textcolor{red}{\textbf{high, delicate pitch}}... her voice carries a \textcolor{red}{\textbf{light, airy}} edge... The texture is \textcolor{red}{\textbf{smooth and clear}}... \\ 
\midrule

\multirow{2}{*}{\textbf{C}} & \textit{GT Caption} & A female speaker \textbf{delivers} the line "..." with a \textbf{lively, upbeat energy} and \textbf{confident presence}. Her voice is \textbf{clear and smooth}, projecting loudly and \textbf{rapidly}, giving her words a sense of \textbf{animated assurance}... \\ \cline{2-3}
& \textit{Sabotaged} & A female speaker \textcolor{red}{\textbf{sings}} the line "..." with a \textcolor{red}{\textbf{melodic, operatic energy}}. Her voice is \textcolor{red}{\textbf{musical and rhythmic}}, projecting loudly \textcolor{red}{\textbf{in a song-like manner}}, giving her words a sense of \textcolor{red}{\textbf{artistic performance}}... \\ 
\midrule

\multirow{2}{*}{\textbf{D}} & \textit{GT Caption} & A \textbf{female} speaker \textbf{delivers} the line "..." with a \textbf{poised, upbeat} tone. Her voice, though \textbf{gentle and subdued} in volume, carries a \textbf{confident, assured} quality. She speaks at a \textbf{measured, unhurried} pace... \\ \cline{2-3}
& \textit{Sabotaged} & A \textcolor{red}{\textbf{male}} speaker \textcolor{red}{\textbf{shouts}} the line "..." with a \textcolor{red}{\textbf{furious, aggressive}} tone. His voice, \textcolor{red}{\textbf{booming and loud}} in volume, carries a \textcolor{red}{\textbf{hostile, threatening}} quality. He speaks at a \textcolor{red}{\textbf{frantic, rushed}} pace... \\ 

\bottomrule
\end{tabularx}
\end{table*}

As observed in Table \ref{tab:case_study}, the generated hallucinations are highly cohesive. For instance, when constructing a Type B hallucination (Attribute Swap), the perturbation does not merely flip the demographic label (``male'' to ``female''); it concurrently updates the correlated acoustic metadata (``low, rumbling pitch'' to ``high, delicate pitch''). This internal logical consistency makes these negative samples far more challenging for standard cross-modal retrieval models to distinguish compared to randomly shuffled text.